\documentstyle[prd,aps,preprint]{revtex}
\tightenlines
\begin{document}
\draft
\title
{ Chiral perturbation theory analysis of
         the baryon magnetic moments revisited}
\author{
      Loyal Durand\thanks{Electronic address:ldurand@theory2.physics.wisc.edu}
and Phuoc Ha\thanks{Electronic address: phuoc@theory1.physics.wisc.edu}}

\address{Department of Physics, University of Wisconsin-Madison,\\
1150 University Avenue, Madison, WI 53706}

\date{\today}

\maketitle

\begin{abstract}
We reexamine critically the chiral expansion for the baryon magnetic moments
including the contributions from loops which involve intermediate octet
and decuplet baryons. We find that, contrary to some claims, the nonanalytic
loop contributions of orders $m_s^{1/2}$ and $m_s\ln m_s$ are of the same
general size because of large coupling factors for the latter, and that the
decuplet contributions are as large as the octet contributions and must be
included in a consistent calculation. There is no clear evidence of the
convergence of the chiral series. The adequacy of the theory will not be
established until dynamical models are able to calculate the contributions
from the counterterms that largely hide the loop effects in fits to the data.
\end{abstract}


\section{INTRODUCTION}

Chiral perturbation theory (ChPT) has been a very useful approach to the theory
of the low momentum processes involving
mesons and baryons, and has been used in various attempts to
explain the baryon magnetic moments. All the moments in the baryon
octet except that of the $\Sigma^0$ have been measured. The $\Sigma^0\Lambda$
transition moment $\mu_{\Sigma\Lambda }$ is also known. At the SU(3)-symmetric
tree level \cite{Coleman}, ChPT parameterizes these eight measured quantities
terms of the two parameters $\mu_D$ and $\mu_F$ in the effective Lagrangian
\begin{equation}\label{lagt}
{\cal L} = {e \over {4 m_N}} ( \mu_D {\rm Tr} \, \bar B \sigma_{\mu \nu}
       F^{\mu \nu} \{ Q, B \} + \mu_F {\rm Tr} \, \bar B \sigma_{\mu \nu}
       F^{\mu \nu} [ Q, B ] ) \ ,
\end{equation}
and is able to fit them with an average error of about 0.24 nuclear magnetons
$\mu_{\rm N}$.
In this expression, $B$ is the usual representation for
the baryon-octet in flavor-space, $Q = {\rm diag}(2, -1, -1)/3 $ is the quark
matrix, $m_N$ is the nucleon mass, and $F^{\mu \nu}$ is the electromagnetic
field.

The study of the baryon moments beyond the tree level has a long history which
can be found in
\cite{Caldi,Gasser,Krause,Jenetal,Dashen,JenkinsManohar,Luty,Daietal,Mei,Bos}.
The approaches and conclusions have varied.
Caldi and Pagels \cite{Caldi} pointed out that the loop corrections lead to
uniquely defined ``nonanalytic corrections'' to the moments proportional
to the square root and logarithm of the symmetry breaking parameter
$m_{\rm s}$ of ChPT. In contrast, the analytic corrections involving integer
powers of $m_{\rm s}$ are cutoff dependent in perturbation theory.
The contributions left after the divergences are subtracted are not uniquely
defined in the absence of further dynamical information, and mix
with the contributions of higher chiral couplings. Since these
contributions are expected to be small, they have generally been ignored.
Caldi and Pagels \cite{Caldi} calculated what were supposed to be the
leading corrections to the baryon moments, those of order
$m_{\rm s}^{1/2}$, and found that they are as large as the tree-level
moments and fail to improve the fit to experiment. These terms were
recalculated by Gasser {\em et al.} \cite{Gasser}, and the logarithmic terms
of type $m_{\rm s}$ln$m_{\rm s}$ were analyzed by Krause \cite{Krause},
but without detailed analyses of the moment problem.

The problem has been reexamined in recent years. Jenkins {\em et al.}
\cite{Jenetal} used heavy baryon chiral perturbation theory (HBChPT)
\cite{HBChPT} to calculate the one-loop corrections to the
octet baryon moments allowing both octet and decuplet intermediate states.
Their formal analysis included the nonanalytic corrections of types
$m_{\rm s}^{1/2}$ and $m_{\rm s}$ln$m_{\rm s}$
and all the counterterms allowed to that order. However, they assumed in
fitting the moment data that the logarithmic corrections and the counterterms
were small compared to the terms proportional to $m_{\rm s}^{1/2}$, and
included only the last in their analysis. The resulting fit
to the data was much worse than the tree-level fit if they used
tree-level values for the axial coupling constants, but better
if they used the axial coupling constants extracted from one-loop  fits
\cite{DFvalues} to other data, $F=0.4$, $D=0.61$, and ${\cal C}=-1.2$,
values about 20\% smaller than those derived at tree level. This
sensitivity leaves the significance of the loop corrections unclear.
In contrast, Meissner and Steininger \cite{Mei}
considered all octet terms (analytic and nonanalytic) which are of order $q^4$
or less in ChPT. They did not include
the intermediate decuplet states directly, but
argued that some decuplet contributions are included because of the way
they determined the couplings. With the counterterms included, they could fit
the seven well measured octet moments precisely, and predicted a
$\Sigma\Lambda$ transition moment of $(1.40\pm 0.01\,\mu_{\rm N}$)
to be compared with the experimental value,
$\mu_{\Sigma\Lambda}=(1.610\pm 0.080)\,\mu_{\rm N}$.
Their principal conclusion was that the chiral expansion appeared to
be converging rather well.

The problem has also been studied using the $1/N_{\rm c}$ expansion by
a number of authors \cite{Dashen,JenkinsManohar,Luty,Daietal} who
showed that the $m_{\rm s}^{1/2}$ terms are leading order in the expansion,
and obtained a number of interesting sum rules for the moments.
Dai {\em et al.} \cite{Daietal} again kept only the leading nonanalytic
corrections proportional to $m_{\rm s}^{1/2}$ and the counterterms
in their analysis of the moment data, and obtained a good overall fit,
but with seven parameters.  Finally, Bos {\em et al.} \cite{Bos} considered the
moments from the point of view of flavor SU(3) breaking in the baryon octet
using a chiral counting and treatment of meson masses different from that
usually used and again obtained a successful parametrization
using seven parameters to describe eight measured moments.

\section{PREDICTIONS OF THE CHIRAL EXPANSION}

Thus far, the numerically successful approaches to the
baryon moments have all
used seven parameters to fit eight experimental data. Two of these parameters
are the tree-level moment parameters $\mu_D\equiv a_1$ and $\mu_F\equiv a_2$.
The remaining five  are the coefficients $a_3,\ldots,a_7$
of the SU(3) symmetry breaking counterterms defined as
\footnote{
We follow the notation used in \cite{Jenetal} except for a rearrangement
of the counterterms following Bos {\em et al.} \cite{Bos}.
}
\begin{eqnarray}
\label{counttt}
     {\cal L}_{ct} &=& {e \over 4m_N}F^{\mu\nu}(
     a_3 {\rm Tr}\,\bar{B}  [[Q,\sigma_{\mu\nu}B], {\cal M}]\,
     + a_4 {\rm Tr}\,\bar{B}  \{[Q,\sigma_{\mu\nu}B], {\cal M}\}\,
    \nonumber\\
   &&+ a_5 {\rm Tr}\bar{B} [\{Q,\sigma_{\mu\nu}B\}, {\cal M}]
     + a_6 {\rm Tr}\bar{B}\{\{Q,\sigma_{\mu\nu}B\}, {\cal M}\}\\
   &&+a_7 {\rm Tr}\bar{B} \sigma_{\mu\nu}B \,{\rm Tr}{\cal M} Q
     + \, ...)\;.
   \nonumber
\end{eqnarray}
Here ${\cal M} = {\rm diag}(0,0,1)$ is a matrix used to introduce SU(3)
breaking introduced by the strange quark mass. A factor $m_{\rm s}$ has been
absorbed in the coefficients of the Lagrangian.
There are two terms with the same structures as those in the tree level moment
operator (\ref{lagt}) hidden in the ellipsis .

Given the large number of parameters used in the fits, it is natural
to ask if the loop corrections matter, and if
one can show that the uniquely extractable nonanalytic corrections actually
improve the tree-level fit to the baryon moments.
To answer these questions and investigate the expected cancellations
and the convergence of the chiral
expansion, we have redone the calculations of Jenkins
{\it et al.} \cite{Jenetal},  corrected several errors in the published work,
\footnote
{
There is a factor $(-5)$ missing in a contribution from the graph 2c and
a factor $1/2$ missing from the decuplet magnetic moment operator. Other
corrections are noted in the {\em erratum} to \cite{Jenetal}.
}
and reanalyzed the problem of fitting the measured moments from a different
point of view. We have followed Jenkins {\em et al.}\ in retaining only the
nonanalytic contributions of types $m_{\rm s}^{1/2}$ and
$m_{\rm s}$ln$m_{\rm s}$ in our analysis, but including the effects of the
octet-decuplet mass splitting throughout.
The graphs which produce these contributions are shown in Figs.\ 1
and 2, respectively.

The octet baryon moments are given in the one-loop approximation by
\begin{equation}
\label{loopsum}
\mu_i = \mu_i^{(0)} + \mu_i^{(1/2)} + \mu_i^{(3/2)} \ ,
\end{equation}
where $\mu_i^{(0)}$ is the tree-level moment, $\mu_i^{(1/2)}$ contains the
contributions from loops that involve only intermediate spin-1/2
octet baryons, and $\mu_i^{(3/2)}$ contains the loop contributions that
involve the spin-3/2 decuplet baryons,
\begin{eqnarray}
\label{loopcontributions}
\mu_i^{(0)}&=&\alpha_i  \ , \\
\mu_i^{(1/2)} &=& \sum_{X=\pi,K} { m_N M_X \over {8\pi f^2}}
        \beta_i^{(X)}  \, + \nonumber \\
       & & \sum_{X=\pi,K,\eta}{1 \over {32\pi^2 f^2}}
         (\gamma_i^{(X)}-2\lambda_i^{(X)}\alpha_i )
	 M_X^2 {\rm ln} M_X^2  \ ,  \\
\mu_i^{(3/2)} &= & \sum_{X=\pi,K} {m_N \over 8\pi f^2} \,
       F(M_X , \delta, \mu) \beta_i^{'(X)} \, + \nonumber \\
     & & \sum_{X=\pi,K,\eta}{1 \over 32\pi^2 f^2}
  \Big [ \ ( \tilde \gamma_i^{(X)}
  -2 \tilde \lambda_i^{(X)}\alpha_i ) L_{(3/2)}(M_X , \delta, \mu) \ +
    \hat \gamma_i^{(X)} L_{(3/2)}^{'}(M_X , \delta, \mu) \Big ]  \, .
\end{eqnarray}
The functions $F$, $L$, and $G$ are defined as
\begin{eqnarray}
\label{FGL}
 \pi \, F (M , \delta, \mu) &=& -\delta \, {\rm ln} {M^2 \over \mu^2} \, + \,
	\left\{
    \begin{array}{ll}
        2\sqrt{M^2-\delta^2} \, [ \pi/2 - {\rm arctan} \,
          (\delta / \sqrt{M^2-\delta^2})], & M \geq \delta ,\\
	\\
         - 2\sqrt{\delta^2-M^2} \, {\rm ln} \,
      [(\delta +\sqrt{\delta^2-M^2}) / M], & M < \delta,
    \end{array}
 \right.,  \\
L_{(3/2)}(M, \delta, \mu) &=& M^2 \, {\rm ln} {M^2 \over \mu^2} \, + \,
	2\pi\delta \, F(M , \delta, \mu),  \\
L_{(3/2)}^{'}(M, \delta, \mu) &=& M^2 \, {\rm ln} {M^2 \over \mu^2} \, + \,
	{2\pi \over 3\delta} \, G(M , \delta, \mu),  \\
\pi \, G (M , \delta, \mu) &=& -\delta^3 \, {\rm ln} {M^2 \over \mu^2} \, + \,
	\pi M^3 \nonumber \\
	&& + \left\{
    \begin{array}{ll}
        2(M^2-\delta^2)^{3/2} \, [ \pi/2 - {\rm arctan} \,
          (\delta / \sqrt{M^2-\delta^2})], & M \geq \delta, \\
	\\
         - 2(\delta^2-M^2)^{3/2} \, {\rm ln} \,
      [(\delta +\sqrt{\delta^2-M^2}) / M], & M < \delta,\label{G}
    \end{array}
 \right. .
\end{eqnarray}
The coupling coefficients $\alpha_i$, $\beta_i$, $\beta_i^{'}$,
$\lambda_i +\tilde \lambda_i$, $ \gamma_i+ \tilde \gamma_i + \hat \gamma_i$ are
identical to those in \cite{Jenetal} where $\lambda_i+\tilde \lambda_i$ is
denoted by $\bar \lambda_i$ and $ \gamma_i+ \tilde \gamma_i + \hat \gamma_i$ is
denoted by $\bar \gamma_i$. The necessary corrections have been included.
The counterterm contributions follow from Eq.\ (\ref{counttt}) and are given in
Table I.

Before examining the loop contributions in detail, we note that
the seven well-measured octet moments can be fitted exactly using
the five counterterm couplings $a_3,\ldots,a_7$ and the two tree-level
parameters $\mu_D\equiv a_1$ and $\mu_F\equiv a_2$, even in the presence of the
loop corrections.\footnote
{
We choose to focus on the seven baryon moments
in our analysis because of the high accuracy with which
they are known relative to $\mu_{\Sigma\Lambda}$. The
alternative of making a weighted least-squares fit to all eight measured
quantities leads to essentially identical results because of the low  weight
accorded $\mu_{\Sigma\Lambda}$ in the fit. The unweighted fits used
in \cite{Daietal,Bos,Mei} distribute the uncertainty in $\mu_{\Sigma\Lambda}$
over all the moments without changing the one test of the theory, as
will be seen below. We believe that the present procedure is clearer.
}
To see this,
we define an $8\times 7$ matrix $X$ which has as its $i$th column the
matrix elements of the corresponding operator in the physical baryon states
$p,\,n,\,\Sigma^+,\,\Sigma^-,\,\Xi^0,\,\Xi^-,\,\Lambda$ and the
$\Sigma^0\Lambda$ transition matrix element as in Table I.
The most general column vector of magnetic moments consistent with this
chiral structure is then of the form
\begin{equation}
\mu=Xa, \label{mu=Xa}
\end{equation}
where $a$ is a seven-component column vector of coefficients.
The $7\times 7$ matrix $\hat{X}$ obtained by omitting the $\Sigma^0\Lambda$
row in $X$ is invertible, so coefficients $a_i$ can be found which give exact
fits to either the seven well-measured baryon moments,
$a=\hat{X}^{-1}\hat{\mu}_{\rm meas}$, or to the difference between
the measured moments and the loop corrections in this sector,
$a'=\hat{X}^{-1}(\hat{\mu}_{\rm meas}-\hat{\mu}_{\rm loop})$. In the
second case, the moments are again reproduced exactly when the loop
corrections are added in.
The only remaining constraint on the theory is
that provided by the $\Sigma^0\Lambda$ transition moment.

In the absence of loop corrections, $\mu_{\Sigma\Lambda}$ is determined
entirely by the seven coefficients $a$ through the product $Xa$ with the full
matrix $X$, or, equivalently,
through the Okubo sum rule \cite{Okubo}
\begin{equation}
4\sqrt{3}\mu_{\Sigma\Lambda}=6\mu_\Lambda + \mu_{\Sigma^-} + \mu_{\Sigma^+}
-4\mu_n - 4\mu_{\Xi^0}. \label{sumrule}
\end{equation}
The origin of the sum rule can be seen rather simply in the present context.
We can extend the matrix $X$ to an $8\times 8$ matrix $\hat{X}$
by adding a final column with zero entries without affecting the seven
baryon moments. That is, $\mu=\hat{X}a$  with the same coefficients $a$
as determined above. However, the extended
matrix has one zero eigenvalue. The sum rule is just the inner product of
the left null eigenvector $\tilde{x}_0$ with $\mu=Xa$, $\tilde{x}_0\mu=
\tilde{x}_0\hat{X}a=0$. In particular, any set of moments which can be
described by the initial seven-component chiral structure must satisfy
the sum rule exactly.

The sum rule is, in fact, satisfied rather accurately by the measured moments.
The predicted value of $\mu_{\Sigma\Lambda}$ is
$\mu_{\Sigma\Lambda}=(1.483 \pm 0.012)\,\mu_{\rm N}$, to be compared with the
measured value $|\mu_{\Sigma\Lambda}|=1.610\pm 0.080)\,\mu_{\rm N}$. The
discrepancy is at the level of 1.5$\sigma$ or 8\% of the measured transition
moment.

The situation is potentially different when loop corrections are
included. While the $m_{\rm s}^{1/2}$ corrections can be
described using only the chiral
structures considered so far and therefore satisfy the Okubo sum
rule exactly \cite{Jenetal}, that is not the case for
the corrections proportional to $m_{\rm s}$ln$m_{\rm s}$, a point
misstated by those authors. The logarithmic corrections involve sums
of terms of the form $M_i^2\ln (M_i/\lambda)$ multiplied by coupling
coefficients. If the mass dependence of the logarithms could be neglected,
the sum rule would hold exactly for external meson masses that satisfy the
Gell-Mann--Okubo mass formula. As a result, the chiral cutoff $\lambda$ drops
out of the sum rule, and the violation depends only on the logarithm of the
mass ratio, $\ln (M_K/M_\pi)$, hence would vanish for equal meson masses.
However, the sum rule is violated for $M_K\not=M_\pi$, and the logarithmic loop
corrections change the prediction for $\mu_{\Sigma\Lambda}$ and introduce
new structure in the $8\times 8$ moment space.\footnote{
The new structure can be parametrized by a single invariant ${\rm Tr}\,\bar{B}
MQBM$ of order $m_s^2$ which affects only the $\Lambda$ moment.}

To investigate this point in more detail, we have constructed
exact fits to the seven octet moments using
the results for the nonanalytic parts of the loop corrections
given in Eqs.\ (\ref{loopsum}-\ref{G}) using the
corrected coupling coefficients from  \cite{Jenetal},
and have then calculated $\mu_{\Sigma\Lambda}$.
For illustration, we will present the results for the two cases considered
in \cite{Jenetal}: (a) $F=0.4$, $D=0.61$, and ${\cal C}=-1.2$,
the one-loop axial vector couplings derived in \cite{DFvalues};
and (b) $F=0.5$, $D=0.75$, and ${\cal C}=-1.5$, the tree-level couplings.
In both cases, we take the decuplet-octet mass difference as $\delta=0.3$ GeV,
and use $f_\pi=0.093$ GeV, $f_K=f_\eta=1.2 f_\pi$, $\mu_T=-7.7$,
$\mu_C=1.94$, and a chiral cutoff $\lambda=1$ GeV as in that reference.

Although the sum rule is violated analytically, we find that
the violation is remarkably small numerically. In particular,
$\mu_{\Sigma\Lambda}$ predicted to be 1.508\,$\mu_{\rm N}$ for the
parameters in case (a), rather the 1.483\,$\mu_{\rm N}$ as required
by the sum rule. The difference is quite small compared to the experimental
uncertainty in $\mu_{\Sigma\Lambda}$. The results in case (b) are similar,
with a predicted transition moment $\mu_{\Sigma\Lambda}=1.534$.
The chiral parameters for these two fits are given in Table II.

It is evident from Table II that the fitted
parameters $\mu_F$ and $\mu_D$ are consistent with SU(6): $\mu_D$/$\mu_F$
$\sim$  3/2. That is expected in the usual quark model picture.
The axial coupling constants $F$, $D$, and ${\cal C}$ also
satisfy the SU(6) relations.

The result that loop corrections lead to only small violations of the Okubo
sum rule appears to be quite robust. In particular, the changes in the loop
corrections associated with different treatments of the meson-baryon couplings
are largely absorbed by changes in the fitted counterterm parameters.
We conclude that the theory with all the counterterms present is
not usefully predictive at the present time. The results do not, for example,
distinguish in a useful way between the standard chiral approach with the
nonanalytic contributions singled out,
the approach of Meissner and Steininger \cite{Mei} which includes analytic
as well as nonanalytic terms, or the approach of Bos {\em et al.}
\cite{Bos} which allows only corrections that break SU(3) symmetry linearly,
fits the data with just the seven parameters above, and satisfies the sum rule
exactly.

\section{ANALYSIS OF THE LOOP CORRECTIONS}

The insensitivity of the overall fits to loop effects raises the question of
whether there is any evidence that these effects are actually important. The
difficulty is that the counterterm contributions are not calculable in
the present theory, yet are essential for getting good overall fits to the
data. A breakdown of the detailed contributions to
the moments for case (b) is given in Table III.
The results for case (a) are similar.

The structure of the loop contributions is of considerable interest. We
will examine this separately for the octet and decuplet cases.

\subsection{Octet baryon contributions}

As shown in Table III, the contributions of type $m_s^{1/2}$ from
the intermediate octet states are quite large, comparable to the tree-level
contributions. The $m_{\rm s} {\rm ln}m_{\rm s}$
corrections have the same general magnitude, but opposite signs.
As a result there are large cancellations
between these two types of nonanalytic contributions. This contradicts the
arguments of chiral perturbation theory \cite{Caldi,Jenetal}
and the $1/N_{\rm c}$ expansion \cite{Dashen,JenkinsManohar,Luty,Daietal}
that the $m_{\rm s} {\rm ln} m_{\rm s}$ corrections should be small, of the
same order of magnitude as the counterterms and
ignorable relative to the $m_{\rm s}^{1/2}$ contributions. In fact, the
smallness of the characteristic ratio \cite{Jenetal}
\begin{equation}
\left(\frac{M_X^2}{32\pi^2f_X^2}\ln\frac{M_X^2}{\lambda^2}\right)
\left(\frac{M_Xm_N}{8\pi f_X^2}\right)^{-1}
=\frac{M_X}{4\pi m_N}\ln\frac{M_X^2}{\lambda^2}
\end{equation}
of the two types of nonanalytic terms, approximately -1/22 and -1/17
for $X$ a pion or a kaon, is offset by very large ratios
of the effective couplings, e.g., 16.1 and 12.7 for the pion and
kaon contributions to the proton moment in case (b), and -28.6 and 9.9
for the $\Xi^0$,
and the large cancellations ensue. There is no indication in this sector
that the chiral perturbation series is converging.

\subsection{Decuplet contributions}

The contributions from intermediate decuplet states introduce a new
complication, namely the dependence of the results on the octet-decuplet
mass difference $\delta$. The results are very sensitive to $\delta$,
which is a QCD parameter rather than a chiral parameter.\footnote{
E.g., $\delta$ is described in the quark model in terms of a color hyperfine
splitting between the spin-1/2 octet and the spin-3/2
decuplet states.}
Meissner and Steininger \cite{Mei} argue that the decuplet contributions
should be included in the chiral momentum expansion, expand in powers
of $q/\delta$, and evaluate their couplings to take the resulting effects
into account. Unfortunately, this argument requires that $\delta$ be
large on the scale of the momenta in the loop integrals. It is
not, with $M_\pi<\delta<M_K$ for typical mass splittings $\delta\approx 300$
MeV, and the heavy decuplet argument fails.  We conclude that the decuplet
states must be regarded as light, on the same mass scale as the octet states,
to obtain a consistent picture. Their contribution to the moments is then
given by the expressions in Eqs. (\ref{loopsum}-\ref{G}). The approximation
in \cite{Mei} becomes useful only for excited baryon multiplets, and can
presumably be used to estimate those contributions.

Jenkins {\em et al.} included the effects of the mass splitting in the
$m_s^{1/2}$ terms, but took $\delta=0$ in the logarithmic contributions. This
introduces very large changes in the latter, but does not affect their final
conclusions because the differences are largely compensated through the
counterterms.

Using the value $\delta=300$ MeV, we obtain the results for the decuplet
contributions shown in Table III. The logarithmic contributions associated
with the decuplet are large
and tend to cancel the logarithmic corrections from the octet as suggested
in \cite{Jenetal}, but the
cancellation is far from complete and all the logarithmic terms must be
included.

If we look at Table III from the point of view of a loop expansion, we see that
the total contributions from the octet loops in Figs.\ 1 and 2 are
considerably smaller than the individual $m_s^{1/2}$ and $m_s\ln m_s$ terms
because of the cancellations. This suggests that a loop expansion may converge
more rapidly than the chiral expansion. However, the two types of terms
tend to add rather than cancel for the decuplet, and turn out to
dominate the final loop corrections. The counterterms are also quite
significant. We see no convincing evidence that the chiral expansion with
only the nonanalytic terms retained is under control. Our attempts to determine
whether or not the loop corrections improve the tree-level fit to the
moments are also inconclusive. We have refit the data omitting the
counterterms, but with the loop corrections included and $\mu_D$ and $\mu_F$
readjusted. The quality of the fits is essentially unchanged with respect
to tree level if all loop contributions are included. Jenkins {\em et al.}
find that the fits can be improved by omitting the logarithmic corrections
and readjusting $F$ and $D$ in addition to the somewhat small values quoted
for case (a), but this procedure is not justified by the relative sizes
of the terms retained and omitted (see Table III).

\subsection{Conclusions}

The foregoing results suggest that the loop
expansion may give a more useful approach to the calculation of the baryon
moments than the chiral and $1/N_{\rm c}$ expansions. In particular, the
results obtained from the different diagrams in Figs.\ 1 and 2 are all of
comparable size when combined
with the relevant coupling factors and must be treated together to get
meaningful results. However, loop effects are largely wiped out in the
fits to the data when all the counterterms
are treated as free parameters. Neither the chiral nor loop expansion is
usefully predictive in this setting given the experimental and theoretical
uncertainties. It is not possible, for example, to distinguish between the
standard chiral expansion and the alternative approach proposed by Bos {\em
et al.} \cite{Bos}. We conclude that it will be necessary to develop a
dynamical theory in which the baryons are treated as composite and the
counterterms become calculable before the moment problem can be regarded
as solved.

\section*{ACKNOWLEDGMENTS}
The authors would like to thank Profs. Elizabeth Jenkins, Roxanne Springer,
and Ulf-G. Meissner for useful correspondence about their published results.
This work was supported in part
by the U.S. Department of Energy under Contract No.
DE-FG02-95ER40896, and in part by the University of Wisconsin Graduate
School with funds granted by the Wisconsin Alumni Research Foundation.

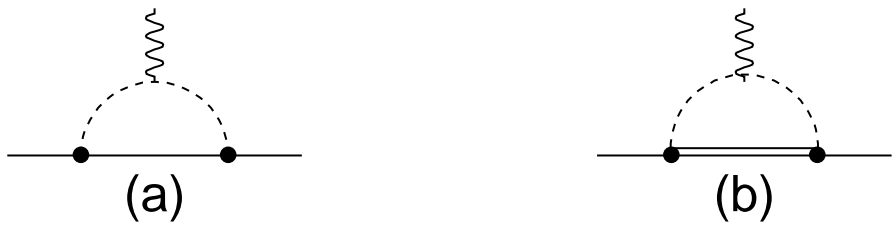
\begin{figure}
\caption{The diagrams which give the non-analytic $m_{\rm s}^{1/2}$
corrections to the baryon magnetic moments. Dashed lines denotes mesons;
single and double solid lines denote octet and decuplet baryons, respectively.
\label{fig:one}}
\end{figure}

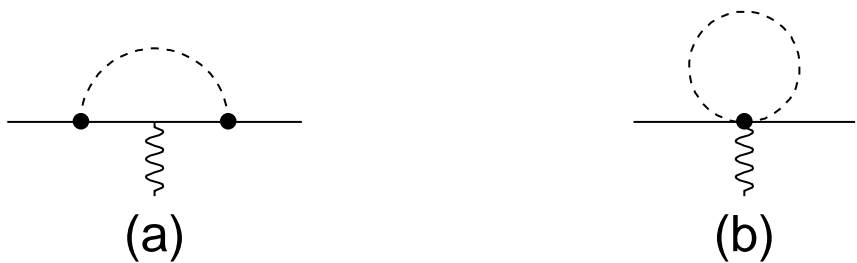
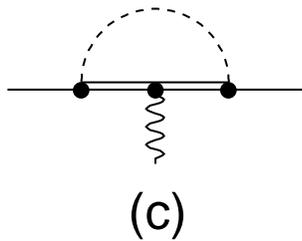
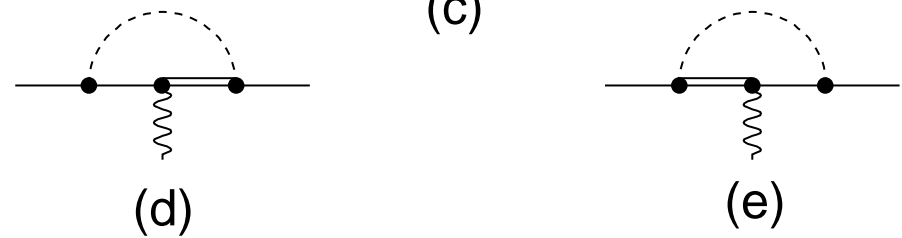
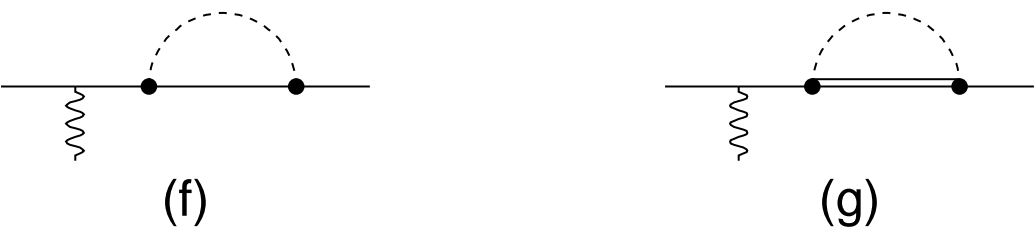
\begin{figure}
\caption{The diagrams which give the non-analytic $m_{\rm s} {\rm ln} m_{\rm
s}$ corrections to the baryon magnetic moments.
\label{fig:two}}
\end{figure}

\newpage


\begin{table}
\caption{Table of the coefficients of the counterterm contributions to
the baryon magnetic moments. This corresponds to the matrix $X$ defined in
the text. The notation is as follows: $a_1=\mu_D$, $a_2=\mu_F$, Eq.\ (1), while
$a_3,\ldots,a_7$ are the
parameters are defined in Eq.\ (2). The $\Sigma^0$
moment is not included, but is given by the SU(2) relation $\mu_{\Sigma^0}=
(\mu_{\Sigma^+}+\mu_{\Sigma^-})/2$.}

\begin{tabular}{cccccccc}
Baryon& $a_1$&$a_2$ &$a_3$ &$a_4$ &$a_5$ &$a_6$ &$a_7$ \\
\hline
$p$&$\frac{1}{3}$ &1 &1 &1 &$\frac{1}{3}$ &$\frac{1}{3}$ & $-\frac{1}{3}$\\
$n$&$-\frac{2}{3}$ &0 &0 &0 &$-\frac{2}{3}$ &$-\frac{2}{3}$ &$-\frac{1}{3}$ \\
$\Sigma^+$&$\frac{1}{3}$ &1 &0 &0 &0 &0 &$-\frac{1}{3}$ \\
$\Sigma^-$&$\frac{1}{3}$ &-1 &0 &0 &0 &0 & $-\frac{1}{3}$\\
$\Xi^0$&$-\frac{2}{3}$ &0 &0 &0 &$\frac{2}{3}$ &$-\frac{2}{3}$ &$-\frac{1}{3}$
\\
$\Xi^-$&$\frac{1}{3}$ &-1 &1 &-1 &$-\frac{1}{3}$ &$\frac{1}{3}$ &$-\frac{1}{3}$
\\
$\Lambda$&$-\frac{1}{3}$ &0 &0 &0 &0 &$-\frac{8}{9}$ &$-\frac{1}{3}$ \\
$\Sigma^0\Lambda$&$\frac{1}{\sqrt{3}}$ &0 &0 &0 &0 &0 &0 \\
\end{tabular}
\end{table}

\begin{table}
 \caption{ The best fit parameters obtained by fitting the seven measured octet
magnetic moments exactly. The parameters are given in units of
$\mu_N$. The $\Sigma\Lambda$ transition moment is predicted to be $\mu_{
\Sigma\Lambda}= 1.508 \mu_{\rm N}$ in case (a), $1.534 \mu_{\rm N}$
in case (b), and $(1.483 \pm 0.012)\,\mu_{\rm N}$ by the Okubo
relation, case (c).}

\begin{tabular}{cddddddd}
        Case  & $a_1$ & $a_2$ &$a_3$ & $a_4$
       & $a_5$ &  $a_6$ & $a_7$  \\
    \hline
    (a)   & 3.183 & 1.992 & 0.269 & $-$0.215 & 0.534 &
              0.224 & $-$0.392 \\
    (b)   & 3.946 & 2.353 & $-$0.001 & $-$0.172 & 0.569 &
              0.694 & $-$1.165 \\
     (c) & 2.568 & 1.809 & 0.591 & $-$0.253 & 0.497 & $-$0.506 & 0.621
\end{tabular}

\end{table}


\begin{table}
 \caption{ Detailed breakdown of the contributions of the loop integrals
and the counterterms to the fitted magnetic moments of the octet baryons
for the parameters of case (b), Table II.
The first seven moments were used as input, and are fitted exactly. The only
prediction of the theory is $\mu_{\Sigma\Lambda}$.}

\begin{tabular}{cdddddddd}
    \hline
    $\mu_B$ & $\mu_D$, $\mu_F$ & $m_s^{1/2}(N)$ & $m_s^{1/2}(\Delta)$
& ln $m_s(N)$ & ln $m_s(\Delta)$ & Loops & CT & $\mu_B$ \\
    \hline
    p           &  3.668 &  $-$2.227  &  $-$0.162 & 2.148 & $-$1.271 & $-$1.512
& 0.636  &2.793 \\
    n           & $-$2.631 &  0.832 & 0.573 & $-$1.491 & 1.258 & 1.172 &
$-$0.454 &$-$1.913 \\
  $\Sigma^+$    & 3.668 &$-$2.843  & $-$0.474 & 2.829 & $-$1.111 & $-$1.598 &
0.388 & 2.458 \\
    $\Sigma^-$  & $-$1.037 & 0.611 & $-$0.349 & $-$0.804 & 0.031 & $-$0.511 &
0.388 & $-$1.160 \\
    $\Sigma^0$  & 1.315 & $-$1.116  & $-$0.411 & 1.012 & $-$0.540 & $-$1.055 &
0.388  & 0.649  \\
    $\Lambda$   & $-$1.315 & 1.116 & 0.411 & $-$1.211 & 0.614 & 0.931 &
$-$0.228 & $-$0.613 \\
    $\Xi^0$     & $-$2.631 & 2.288 & 0.698 & $-$2.650 & 0.739 & 1.075 & 0.306 &
$-$1.250 \\
  $\Xi^-$    & $-$1.037 & 1.339 & $-$0.286 & $-$1.324 & 0.057 & $-$0.215 &
0.601 & $-$0.651 \\
    $\Sigma^0\Lambda$ &  2.278 & $-$1.157 & $-$0.496 & 1.680 & $-$0.771 &
$-$0.745 & 0.000 & 1.534 \\
    \hline
\end{tabular}

\end{table}


\begin{references}

\bibitem{Coleman}
S. Coleman and S.L. Glashow, Phys.\ Rev.\ Lett.\ {\bf 6}, 423 (1961).
%
\bibitem{Caldi}
D.G. Caldi and H. Pagels, Phys.\ Rev.\ D {\bf 10}, 3739 (1974).
%
\bibitem{Gasser}
J. Gasser, M. Sainio, and A. Svarc, Nucl.\ Phys.\ {\bf B307}, 779 (1988).
%
\bibitem{Krause}
A. Krause, Helv.\ Phys.\ Acta {\bf 63}, 3 (1990).
%
\bibitem{Jenetal}
E. Jenkins, M. Luke, A.V. Manohar and M. Savage, Phys.\ Lett.\ B {\bf 302},
482 (1993);  (E) {\em ibid.} {\bf 388}, 866 (1996).
%
\bibitem{Dashen}
R. Dashen, E. Jenkins, and A.V. Manohar, Phys.\ Rev. D {\bf 49}, 4713 (1994).
%
\bibitem{JenkinsManohar}
E. Jenkins and A.V. Manohar, Phys.\ Lett.\  B {\bf 335}, 452 (1994).
%
\bibitem{Luty} M.A. Luty, J. March-Russell, and M. White, Phys.\ Rev.\ D
{\bf 51}, 2332 (1995).
%
\bibitem{Daietal}
J. Dai, R. Dashen, E. Jenkins, and A.V. Manohar, Phys.\ Rev.\ D {\bf 53},
273 (1996).
%
\bibitem{Mei}
Ulf-G. Mei{\ss}ner, S. Steininger, Nucl.\ Phys.\ {\bf B499}, 349 (1997).
%
\bibitem{Bos}
J.W. Bos, D. Chang, S.C. Lee, Y.C. Lin and H.H. Shih, Chin. J. Phys. (Taipei)
{\bf 35}, 150 (1997).
%
\bibitem{HBChPT}
H. Georgi, Phys.\ Lett.\ B {\bf 240}, 447 (1990);
E. Jenkins and A.V. Manohar, Phys.\ Lett.\ B {\bf 255}, 558 (1991); UCSD/PTH
91-30.
%
\bibitem{Okubo}
S. Okubo, Phys.\ Lett.\ {\bf 4}, 14 (1963).
%
\bibitem{DFvalues}
E. Jenkins and A.V. Manohar, in {\em  Proc.\ Workshop on Effective Field
Theories of the Standard Model}, edited by U. Meissner (World Scientific,
Singapore, 1992); M.N. Butler, M.J. Savage, and R.P. Springer, Nucl.\ Phys.\
{\bf B399}, 69 (1993).

\end{references}
\end{document}